\input harvmac
\input epsf.tex
\noblackbox
\overfullrule=0pt
\def\Title#1#2{\rightline{#1}\ifx\answ\bigans\nopagenumbers\pageno0\vskip1in
\else\pageno1\vskip.8in\fi \centerline{\titlefont #2}\vskip .5in}

%
%
%

\def\[{\left [}
\def\]{\right ]}
\def\({\left (}
\def\){\right )}

\def\p{\partial}

\font\cmss=cmss10 \font\cmsss=cmss10 at 7pt
\def\IZ{\relax\ifmmode\mathchoice
   {\hbox{\cmss Z\kern-.4em Z}}{\hbox{\cmss Z\kern-.4em Z}}
   {\lower.9pt\hbox{\cmsss Z\kern-.4em Z}}
   {\lower1.2pt\hbox{\cmsss Z\kern-.4em Z}}\else{\cmss Z\kern-.4emZ}\fi}

%

\lref\arl{H. Araki and E. Lieb, Comm. Math. Phys. {\bf 18} (1970) 160.}
\lref\dasmathurtwo{S. Das and S. Mathur,
``{\it Interactions Involving D-branes}'',
hep-th/9607149.}

\lref\gk{ S. Gubser and I. Klebanov, 
``{\it Emission of Charged
Particles
from Four- and Five-dimensional Black Holes}'', 
hep-th/9608108.}

\lref\page{ D. Page, Phys. Rev. {\bf D 13} (1976) 198; Phys. Rev. {\bf
D14}
(1976), 3260.}

\lref\unruh{W. Unruh, 
Phys. Rev. {\bf D14} (1976) 3251.}
\lref\cj{ M. Cvetic and D. Youm, hep-th/9512127.}

\lref\tseytlin{A. Tseytlin, hep-th/9601119.}
\lref\tata{S. Dhar, G. Mandal and S. Wadia, ``{\it
Absorption vs. Decay of Black Holes in String Theory and 
T-symmetry}'', hep-th/9605234.}
\lref\gm{G. Horowitz and D. Marolf, hep-th/9605224.}
\lref\mss{J. Maldacena and L. Susskind, ``{\it D-branes and Fat 
Black Holes }'', Nucl. Phys. {\bf B475} (1996) 679,  hep-th/9604042.}
\lref\bl{V. Balasubramanian and F. Larsen, hep-th/9604189.}
\lref\jmn{J. Maldacena,``{\it Statistical Entropy of Near Extremal
Fivebranes}'', hep-th/9605016.}
\lref\hlm{G. Horowitz, D. Lowe and J. Maldacena, ``{\it Statistical
Entropy of Nonextremal Four Dimensional Black Holes and U-duality}'',
Phys. Rev. Lett. {\bf 77} (1996) 430, hep-th/9603195.}
\lref\ms{J. Maldacena and A. Strominger, Phys. Rev.  Lett. {\bf 77}
(1996) 428,  hep-th/9603060.}
\lref\bd{See N. D. Birrel and P.C. Davies,``{\it Quantum Fields in
Curved
Space}'', Cambridge University Press 1982.}
\lref\hms{G. Horowitz, J. Maldacena and A. Strominger,
``{\it Nonextremal Black Hole Microstates and U-duality}'', 
 Phys. Lett. {\bf B383} (1996) 151,
 hep-th/9603109.}
\lref\dbr{J. Polchinski, S. Chaudhuri, and C. Johnson, hep-th/9602052.}

\lref\jp{J. Polchinski, 
``{\it Dirichlet Branes and Ramond-Ramond
Charges}'', Phys. Rev. Lett. {\bf 75}
(1995) 4724, hep-th/9510017.}

\lref\dm{S. Das and S. Mathur, ``{\it
Comparing Decay Rates for Black Holes and D-branes}'',hep-th/9606185;
``{\it Interactions Involving D-branes}'',
hep-th/9607149.}
\lref\ghas{G. Horowitz and A. Strominger,``{\it Counting States of
Near-extremal Black Holes }'', Phys. Rev. Lett. {\bf 77} (1996) 2368,
  hep-th/9602051.}
\lref\ascv{A. Strominger and C. Vafa, ``{\it On the Microscopic 
Origin of the Bekenstein-Hawking Entropy}'', Phys. Lett. {\bf B379} (1996)
99,
 hep-th/9601029.}
\lref\spin{
 J.C. Breckenridge, R.C. Myers, A.W. Peet and  C. Vafa, { \it
D-Branes and Spinning Black Holes}'',  hep-th/9602065.}
\lref\clifford{C. Johnson, R. Khuri and R. Myers, {\it 
Entropy of 4D Extremal Black Holes}'', Phys. Lett. {\bf B378} (1996) 78,
 hep-th/9603061.}

\lref\hrva{P. Horava, Phys. Lett. {\bf B231} (1989) 251.}

\lref\cakl{C. Callan and I. Klebanov, ``{\it D-brane Boundary State
Dynamics}'',
Nucl .Phys. {\bf B465 }(1996) 473, hep-th/9511173.}

\lref\prskll{J. Preskill, P. Schwarz, A. Shapere, S. Trivedi and
F. Wilczek, Mod. Phys. Lett. {\bf A6} (1991) 2353. }
\lref\bhole{G. Horowitz and A. Strominger,
Nucl. Phys. {\bf B360} (1991) 197.}
\lref\bekb{J. Bekenstein, Phys. Rev {\bf D12} (1975) 3077.}
\lref\hawkirr{S. Hawking, Phys. Rev {\bf D13} (1976) 191.}
\lref\stas{A.~Strominger and S.~Trivedi,  Phys.~Rev. {\bf D48}
 (1993) 5778.}
\lref\bek{J. Bekenstein, Lett. Nuov. Cimento {\bf 4} (1972) 737,
Phys. Rev. {\bf D7} (1973) 2333, Phys. Rev. {\bf D9} (1974) 3292.}
\lref\hawkb{S. Hawking, Nature {\bf 248} (1974) 30, Comm. Math. Phys.
{\bf 43} (1975) 199.}
\lref\cama{C. Callan and J. Maldacena, ``{\it The D-brane approach to 
black hole quantum mechanics}'', Nucl. Phys. {\bf B 475 } (1996)
645, hep-th/9602043.}
\lref\hpc{S. Hawking, private communication.}
\lref\bdpss{T. Banks, M. Douglas, J. Polchinski, 
S. Shenker and A. Strominger, in progress.}
\lref\ast{A. Strominger, {\it Statistical Hair on Black Holes''},
hep-th/9606016.  }
\lref\ka{ B. Kol and A. Rajaraman, ``{\it
Fixed Scalars and Suppression of Hawking Evaporation}'',  hep-th/9608126. }

\lref\thooft{G. 't Hooft, 
``{\it The Scattering Matrix Approach for the Quantum Black Hole, 
an Overview}'', gr-qc/9607022.}

\lref\vafainst{ M. Bershadsky, V. Sadov and  C. Vafa, ``{\it D-Strings on
D-Manifolds}'', Nucl. Phys. {\bf B463} (1996) 398,  hep-th/9511222;
C.  Vafa, ``{\it Instantons on D-Branes}'',  Nucl. Phys. {\bf B463} (1996) 435,
hep-th/9512078.}

\lref\becspin{  J. C. Breckenridge, D. A. Lowe, 
R. C. Myers, A. W. Peet, A. Strominger and  C. Vafa, ``{\it
Macroscopic and Microscopic Entropy of Near-Extremal Spinning Black
Holes}'', Phys. Lett. {\bf B381} (1996) 423,  hep-th/9603078.}

\lref\jmas{J. Maldacena and A. Strominger, ``{\it Black Hole Greybody
Factor and D-brane Spectroscopy}'', hep-th/9609026.}

\lref\cgkt{C. Callan, S. Gubser,  I. Klebanov and A. Tseytlin, 
``{\it Absorption
of Fixed Scalars and the D-Brane Approach to Black Holes}'',
 Nucl .Phys. {\bf B489} (1997) 65, hep-th/9610172.}

\lref\verlindemoore{R. Dijkgraaf, G, Moore, E. Verlinde and H. Verlinde, 
``{\it Elliptic Genera of Symmetric Products and Second Quantized
Strings}'',
hep-th/9608096.}

\lref\verlindecount{ R. Dijkgraaf, E. Verlinde and H. Verlinde, ``{\it
Counting Dyons in N=4 String Theory}'',
 hep-th/9607026.}

\lref\nocouplings{ B. de Wit, P. Lauwers and A. Van Proeyen, 
``{\it Lagrangians of N=2 Supergravity Matter Systems''}, Nucl. Phys.
{\bf B255} (1985) 569.}

\lref\seibergho{N. Seiberg, ``{\it Naturalness vs. Supersymmetric 
Non-renormalization Theorems}'', Phys. Lett. {\bf B318} (1993) 469,
 hep-th/9309335; ``{\it Power of Holomorphy- Exact Results 
in 4D Supersymmetric
Field Theories}'', PASCOS 1994, 357-369 (QCD161:I69:1994), hep-th/9408013.}

\lref\seibergthree{N. Seiberg and E. Witten, 
``{\it Gauge Dynamics and Compactification to Three Dimensions}'', 
hep-th/9607163.}

\lref\hawkingunitarity{ S. W. Hawking, {\it Breakdown of
Predictability
in Gravitational Collapse}'', Phys. Rev. {\bf D14} (1976) 2460.}

\lref\douglasfivebrane{ M. Douglas, ``{\it Gauge Fields and Fivebranes}'', 
hep-th/9604198.}

\lref\metricfive{T. Banks, M. Douglas, J. Polchinski, S. Shenker and 
A. Strominger, private comunication.}

\lref\douglas{ M. Douglas, {\it Branes within Branes},
hep-th/9512077.}

\lref\polchinski{
J. Polchinski, ``{\it Monopole Catalysis: The Fermion Rotor System}'',
Nucl. Phys. {\bf B242} (1984) 345; Rubakov, Nucl. Phys. {\bf B203}
(1982)
311; C. Callan, Phys. Rev. {\bf D25} (1982) 2141; {\bf D26} (1982)
2058;
Nucl. Phys. {\bf B212} (1983) 391.}

\lref\jmthesis{J. Maldacena, ``{\it Black Holes in String Theory}'',
Ph.D. thesis, Princeton University 1996, hep-th/9608235.}

\lref\hawkingbck{Reference on the area law.}

\lref\notesondbranes{J. Polchinski, S. Chaudhuri and  C. Johnson,
{\it Notes on D-Branes}, hep-th/9602052.}

\lref\sm{ J. Maldacena and A. Strominger, ``{\it
Statistical Entropy of Four-Dimensional Extremal Black Holes}'',
Phys. Rev. Lett. {\bf 77} (1996) 428, 
 hep-th/9603060.}

\lref\ktmbranes{ I. Klebanov and A. Tseytlin, ``{\it Intersecting M-Branes
as Four Dimensional Black Holes}'',
Nucl. Phys. {\bf B475} (1996) 179,
 hep-th/9604166; V. Balasubramanian and F. Larsen, 
``{\it On D-Branes and Black Holes in Four Dimensions}'', hep-th/9604189. }  

\lref\gkfour{ S. Gubser and I. Klebanov, ``{\it 
Four Dimensional Grey Body Factors and the Effective String}'', 
hep-th/9609076.} 

\lref\proyen{
B. de Wit, P. Lauwers and A. Van Proeyen, ``{\it Lagrangians of
N=2 Supergravity-Matter systems}'', Nucl. Phys. {\bf B255} (1983) 569.}

\lref\hawkingrad{
S. Hawking, ``{\it Particle Creation by Black Holes}'',
 Comunn. Math. Phys. {\bf 43} (1975) 199.}

\lref\complementarity{ G. 't Hooft, ``{\it The Black 
Hole Interpretation of String theory}'', Nucl.Phys {\bf B 335} (1990)
138;  G. 't Hooft, {\it The Scattering Matrix Approach for the 
Quantum Black Hole: An Overview}'' Int. J. Mod. Phys. {\bf A11}
(1996)4623, gr-qc/9607022;
 Y. Kiem, H. Verlinde
and E. Verlinde,
``{\it Black Hole Horizons and Complementarity}'',
 Phys.Rev.{\bf D52}, (1995) 7053 hep-th/9502074.}

\lref\boosts{D. Lowe, J. Polchinski, L. Susskind, 
L. Thorlacius and J. Uglum, 
``{\it  Black Hole Complementarity vs. Locality}'',
Phys.Rev. {\bf D52} (1995) 6997, 
 hep-th/9506138.}

\lref\intrilligator{K. Intrilligator and N. Seiberg, ``{\it Mirror Symmetry in 
Three Dimensional Gauge Theories}'', hep-th/9607207.}

\lref\alvarez{ L. Alvarez-Gaume and D. Freedman, ``{\it 
Geometrical Structure and Ultraviolet Finitness in the Supersymmetric
Sigma Model}'', Comm. Math. Phys. {\bf 80} (1981) 443, and references
therein.}

\lref\dkps{
M. Douglas, D. Kabat, P. Pouliot and S. Shenker, 
``{\it  D-branes and Short Distances in String Theory}'',
hep-th/9608024.}

\lref\bfss{
 T. Banks, W. Fischler, S.  Shenker and  L. Susskind,
``{\it  M Theory As A Matrix Model: A Conjecture}'', hep-th/9610043.} 

\lref\ghr{ S. Gates, C. Hull and M. Ro\v{c}ek, ``{\it
Twisted Multiplets and New Supersymmetric Non-Linear
$\sigma$-Models}'', Nucl. Phys. {\bf B248} (1984) 157.
}

\lref\dps{
M. Douglas, J. Polchinski and A. Strominger, 
``{\it Probing Five Dimensional Black Holes with D-branes}'',
hep-th/9703031.}

\lref\tseytlindbi{A. A. Tseytlin, ``{\it
On non-abelian generalisation of Born-Infeld action in string theory}'',
hep-th/9701125.}

\lref\bachas{ C. Bachas, {\it
D-brane Dynamics}, Phys. Lett. 
{\bf B374} (1996) 37, hep-th/9511043.}

\lref\callancorrections{ C. Callan, C. Lovelace, C. Nappi and S. Yost,
``{\it Loop Corrections to Superstring Equations of Motion}'', Nucl. Phys.
{\bf B308} (1988) 221.}

\lref\callanconstant{ A. Abouelsaood,
C. Callan, C. Nappi and S. Yost,
``{\it Open Strings in Background Gauge 
Fields}'', Nucl. Phys. 
{\bf B280} (1987) 599.}

\lref\douglasseiberg{
T. Banks, M. Douglas, N. Seiberg,
``{\it Probing F Theory with branes}'', Phys. Lett. {\bf B387} (1996)
278, 
hep-th/9605199.}

\lref\lifschytz{ G. Lifschytz, ``{\it Comparing D-branes to Black-branes}'',
Phys. Lett. {\bf B388} (1996) 720, hep-th/9604156.}

\lref\lima{
G. Lifschytz and S. Mathur, ``{\it 
Supersymmetry and Membrane interactions 
in M(atrix) Theory}, hep-th/9612087.}''

\lref\peetsolutions{
A. Peet, ``{\it Entropy and Supersymmetry
of D dimensional Extremal Electric
Black Holes vs. String States}'', 
Nucl. Phys. {\bf B456} (1995) 732,  hep-th/9506200.}

\lref\cvetic{ M. Cvetic, ``{\it Properties of  Black Holes 
in Toroidally Compactified String Theory}'',  hep-th/9701152, and
references
therein.}

\lref\martinec{ M. Li and E. Martinec,
``{\it Matrix Black holes}'', hep-th/9703211;  hep-th/9704134.}

\lref\verlindebh{ 
R. Dijkgraaf, E. Verlinde and H. Verlinde,
``{\it 5-D Black Holes and Matrix Strings}'',
hep-th/9704018.}

\lref\lennyfiniten{
L. Susskind, ``{\it Another Conjecture about
M(atrix) Theory}'',  hep-th/9704080.}

\lref\incredible{
W. Fischler, E. Halyo, A. Rajaraman and
 L.
Susskind, {\it The Incredible Shrinking 
Torus}, hep-th/9703102.}

\lref\oamb{
O. Aharony and M. Berkooz, 
``{\it Membrane Dynamics in M(atrix) Theory}'',
hep-th/9611215.}

\lref\seiberg{ T. Banks, N. Seiberg, private communication.}
\lref\dkpp{
D. Kabat and P. Pouliot,
``{\it  A Comment on Zero-Brane Quantum 
Mechanics}'', Phys. Rev. Lett. 
{\bf 77} (1996) 1004, hepth/9603127.}

\lref\leigh{R. Leigh, ``{\it Dirac Born Infeld action
from Dirichlet Sigma Model}'',
Mod. Phys. Lett. {\bf A4} (1989) 2767.}

\lref\hstro{ G. Horowitz and A. Strominger,
``{\it Black Strings and p-Branes}'',
Nucl. Phys. {\bf B360} (1991) 197. }

\lref\taylor{W. Taylor, ``{\it D-brane Field Theory on Compact 
Spaces}'', 
Phys. Lett. {\bf B394} (1997) 283,
hep-th/9611042.} 

\lref\mdns{M. Dine and N. Seiberg, to appear. }

\lref\tseytinfadeev{E. Fradkin and A. Tseytlin,
``{\it Quantum Properties of higher dimensional and dimensionally 
reduced Supersymmetric Theories}'',  Nucl. Phys. {\bf B198}
(1982) 474.}

\lref\tseytlinmet{R. Metsaev and A. Tseytlin, ``{\it
On Loop Corrections to String Theory Effective Actions}'',
 Nucl. Phys. {\bf B298}
(1988) 109.}

\lref\makeenko{I. Chepelev, Y. Makeenko and K. Zarembo, 
``{\it Properties of D-branes in Matrix Model of IIB Superstring}'',
hep-th/9701151.}

    
%
\Title{\vbox{\baselineskip12pt
\hbox{hep-th/9705053}\hbox{ RU-97-30}}}
{\vbox{
{\centerline { Probing near extremal black holes with D-branes }}
  }}
\centerline{Juan Maldacena\foot{malda@physics.rutgers.edu}}
\vskip.1in
\centerline{\it Department of Physics and Astronomy, Rutgers University,
Piscataway, NJ 08855, USA}
\vskip.1in
\vskip .5in

\centerline{\bf Abstract}

We calculate the one loop effective action
for  D-brane probes moving in the presence of
near BPS D-branes. 
The $v^2$ term agrees  with supergravity in all cases and
the static force agrees for a   five dimensional
black hole with two large charges.
It also agrees qualitatively in all the  other cases. 
We make some comments on  the M(atrix) theory interpretation
of these results.

 \Date{}

%

\newsec{ Introduction }

D-branes are localized probes of the  spacetime  geometry
\douglasfivebrane 
\douglasseiberg \dkpp \dkps .
When a D-brane probe is in the presence of other D-branes there are
massive open strings stretching between the probe and the other
branes. 
When these massive open strings are integrated out one obtains
an effective action for the massless fields representing 
the motion of the probe.
This can be  interpreted  as the action of a D-brane
moving on a nontrivial supergravity background.
In many cases one can find the exact supergravity
backgrounds in this fashion \douglasfivebrane 
\douglasseiberg \dkps .
Most of the backgrounds analyzed previously correspond to BPS supergravity
solutions. 
The calculation in  \dps\ of D-brane probes moving in a 
nonsingular extremal black hole background shows
agreement for the one loop term but the status of the two loop term
is not clear. 

In this paper we study
 probes 
moving in near BPS backgrounds. In the D-brane language this
corresponds to a D-brane probe moving close to 
other D-branes which are not in their ground state but that 
have some additional (small) energy above extremality.

We calculate the one loop effective action for a variety of near 
extremal configurations and we compare it to the corresponding
supergravity results. 
 Since the background  is 
no longer BPS and supersymmetry is broken there is no reason for
the forces to cancel. Indeed there is a net force on a static probe.
We also calculate the $v^2$ force and we find agreement with
supergravity
 in {\it all }
cases.
We compute the static force 
for a  D-brane  configuration with $Q_5$ D-fivebranes carrying also $Q_1$
D-onebrane
charge and some extra energy. We find precise agreement. The static force
agrees qualitatively in all the other  cases.
All one loop calculations of this type reduce to evaluating
 $F^4$ terms in the gauge theory.

\newsec{The one loop calculation}

In this section we will consider a D-brane probe in the presence
of  some other D-branes. By integrating out the stretched 
open strings we  calculate the one-loop effective 
action for the probe.
More concretely, we consider 
$N+1$ D-branes, $N$ of which sit at the origin ($r=0$)  and
the last is  the probe which sits at a distance $r$.
At low energies and for small separations the system is
described by a $p+1$ dimensional 
$U(N+1)$ Yang-Mills theory with 16 supersymmetries
broken down
to $U(N)\times
U(1)$ by the expectation value of an adjoint scalar which 
measures the distance from  the probe to the rest of the branes.
The fields  with one index in $U(N)$ and the other on $U(1)$ 
are massive, with a mass $ m = r/(2 \pi \alpha')$. 
If we integrate them out we get an effective action 
for the light degrees of freedom.
All  D-branes  have $p+1$ worldvolume dimensions, with $p \le 6$.
The action of the Yang Mills theory is 
\eqn\action{ 
S_0 = - { 1 \over g (2\pi)^p } \int d^{1+p}x  \ \ 
{1\over 4} Tr[ F_{\mu\nu} F^{\mu\nu} ]
+~~
{\rm fermions }~,
}
where $\mu,\nu$ are ten dimensional indices \notesondbranes .
We normalize the string coupling $g$ so that $g\to 1/g$ under 
S-duality and set $\alpha' =1$\foot{
These conventions are the ones used in 
in \hms \jmthesis .}. The gauge field is a hermitian matrix and
the trace is defined without any additional coefficients.

We will calculate the leading order terms in the field strength 
$F$.
 We will assume that $F$ is slowly varying so that
we can neglect derivatives,  $DF$, as well as commutators, $[F,F]$. 
Commutators are small if covariant derivatives of the fields
are small since $ [ D , D ] F \sim [ F, F ] $.
A similar  approach was taken in \tseytlindbi\ to propose a form
for the non-abelian generalization of the Dirac-Born-Infeld action.

We consider first the case where the probe D-brane is static.
The one loop contribution to the effective action will 
come from a one loop diagram with four external lines corresponding
to light $U(N)$ fields on the excited D-branes. 
Diagrams with less than four external lines cancel due to
supersymmetry.

\vskip 1cm
\vbox{
{\centerline{\epsfxsize=2in \epsfbox{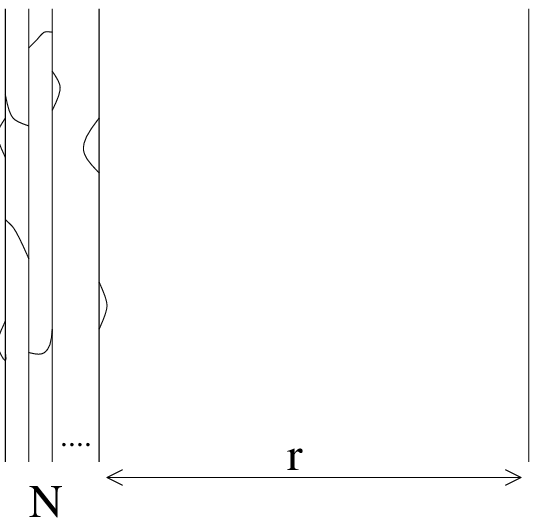}}}
{\centerline{\tenrm FIGURE 1: D-brane configuration. N D-branes
are sitting together carrying }}
{\centerline{ low energy  excitations and the probe is separated 
by a distance $r$.}}
}
\vskip .5cm

Instead of doing the field theory calculation we
do the string theory calculation. It corresponds to a one loop
string diagram with the string ending on the probe on one 
side and the excited brane on the other.
In order to find  the term that we are interested in, it
is enough to  do the calculation for constant field strength $F$.
Since we are going to neglect commutators we can 
take it to be diagonal in the group indices.
This calculation was done in \callancorrections \callanconstant 
\tseytinfadeev \tseytlinmet\
and it is a very simple generalization of the calculations
done by \bachas \cakl \lima \dkps \makeenko . We 
refer to those  papers for the details.

The result is 
\eqn\result{
S_1 = { c_{7-p}\over 8 ( 2\pi )^p  r^{7-p} } 
\int d^{1+p } x \ \left\{ Tr [ F_{\mu \nu} F^{\nu \rho} F_{\rho \sigma }
F^{\sigma \mu }]  - {1 \over 4 } Tr [ F_{\mu \nu}F^{\mu \nu } 
F_{\rho \sigma }F^{\rho \sigma } ] \right\}~,
}
where $\mu , \nu$ are again ten dimensional indices and
$c_q$ is the numerical constant 
\eqn\defcp{
c_q = (4  \pi)^{{q  \over 2 } -1 } \Gamma( { q \over 2 })~ . 
}
We see that there are two terms (remember that we are neglecting
commutators, $[F,F]\sim 0$)
 with a structure that  is independent of the
dimensionality of the brane. This structure is the same
as the structure of the $F^4$ terms in the Dirac-Born-Infeld 
action \tseytlindbi \tseytlinmet .
We have done the calculation for diagonal field strengths but
we wrote \result\ in terms of the full non-abelian $F$, which
is consistent with our approximations.
The result \result\ is correct to all orders in $\alpha'$ for reasons 
explained in \dkps ,
so the gauge theory calculation, which was a priori correct only
for $r \ll \sqrt{\alpha'}$,  also gives the right result for 
 $r \gg \sqrt{\alpha'}$, a fact that  is 
important  in M(atrix) theory \bfss .

The case of branes moving with constant velocity corresponds
to $F_{0i} =v^i$ and  we get \bachas \cakl \dkps  
\eqn\vfourth{
S = { c_{7-p} \over 8  r^{7-p} } R^p \int dt ( v^2 )^2~,
}
where $R$ is the radius of each of the $p$ compact directions. 
This is 
 is the right form for the $v^4$ term in supergravity.

We also see that if $F$ describes a BPS excitation \result\ vanishes.
The two possible BPS excitations are traveling waves (momentum
along the brane) and instantons (for $p \ge 4$) describing
a $p-4 $ brane inside a $p$ brane \douglas .
For a traveling wave we only have 
 $F_{- i} \not = 0$ where $ x^- = t - x^9$, and both terms in
\result\ individually cancel.
For the case of the instanton let us denote by $I,J,K,L$ the
dimensions along which the gauge field is nontrivial (the
four dimensions of the instanton).
Using the self duality condition we find
\eqn\simpleinst{
F_{IJ} F^{JK  } = { 1\over 4 } \delta_{IK} F_{L J} F^{ J L}~,
 }
so that the two terms in \result\ cancel.
We also obtain a cancelation if we have instantons and traveling
waves at the same time.

\subsec{Calculation of the static 
force for the five dimensional  black
hole}

Now we want to excite the $N$ branes and 
evaluate this term in some generic thermal ensemble. 
This is  difficult in principle  because we have not calculated the 
fermionic terms and they will contribute. We will do the calculation
in a case where it is easy to see what the effect of the fermions is.
Of course, as explained in \ascv\  the supergravity solution
is expected to agree only in the limit of large $g N $. Which means
that the effective large $N$ coupling of the gauge theory is strong.

We will calculate \result\ in a configuration carrying the charges 
of the 
the five dimensional near extremal black hole of \cama \ghas\ in the
dilute gas regime. 
Then  $p=5$,  we call $N = Q_5$ and
we also put $Q_1$ instanton strings along one of the directions
of the fivebrane, let us call it the direction $\hat 9$. Even though
they are called ``instantons'' these objects are physically string
solitons
of the 1+5 dimensional gauge theory.
The instanton configuration is characterized by some moduli $\xi^r$,
$r = 1,.., 4Q_1Q_5$.
These moduli can oscillate when we move along the direction $\hat 9$,
$\xi(t,x_9)$. We are interested in the case where the energy of the
oscillations is small, so that we can describe the excitations
of the system as oscillations in moduli space. 
The condition  is that the total energy due to the
oscillations
should be much smaller than the energy of the instantons $ E \ll R_9
Q_1/g $.
(In the notation of \hms \jmas\ this means $r_n \ll r_1 $.)
This picture of the D-1-brane charge being carried by instantons
in the gauge theory is correct when the energy of the instantons
is much smaller that the total  mass of the fivebranes 
$ M_1 = R_9 Q_1/g \ll M_5 =  R_5R_6R_7R_8R_9 Q_5/g $ 
(this  means $r_1 \ll r_5$).
So that we are in the dilute gas regime of \jmas \foot{
The definition of the dilute gas regime $r_n \ll r_1 , r_5 $ does
not require any specific relation between $r_1$ and $r_5 $.
}
Calling $x^\pm = x^9 \pm t$,  the nonzero components of the gauge
field are $F_{\pm I} = \partial_\pm \xi^r \partial_r A_I $ with 
$I = 5,6,7,8$ and $F_{IJ}$,  the field of the instanton.
The action for the small  fluctuations of the instanton configuration
becomes
\eqn\actionfl{\eqalign{
S_0 =& 
{ 1 \over g (2\pi)^5 }
 \int d^{1+5} x  \ { 1\over 2} Tr[ F_{\alpha I } F^{\alpha I} ]
+~~
{\rm fermions } \cr = & { 1\over 2 } \int dt dx^9 
G_{rs}(\xi) \partial_\alpha
\xi^r
 \partial^\alpha \xi^s +~~
{\rm fermions }~,
}}
where $\alpha = \pm $.
We  have an nonlinear sigma model 
action for the instanton fluctuations \ascv . 
The theory \actionfl\ has (4,4) supersymmetry and the metric
$G_{rs}$ is hyperk\"ahler \ascv . 

Using \simpleinst\ 
we can see that the effective action \result\ reduces to 
\eqn\potential{
S_1 = { 2 \over  ( 2\pi )^5  r^{2} }  
\int d^{1+5 } x \  \ Tr [ F_{+I} F_{+ I} F_{-J}
F_{-J }] +~~
{\rm fermions }~.
}
When we integrate over $t,x^9$ we will effectively average 
separately the term with $++$ and the ones proportional to $--$.
We assume that the oscillations average the fields in such a way
that we can replace $ F_{+I} F_{+I} $ by its average value, both 
in spacetime indices and group indices. So we get
\eqn\potentialnew{\eqalign{
S_1 = &{ 2  \over  ( 2\pi )^5  r^{2} } 
\int dt \int d^{5 } x \ Tr [ F_{+I} F_{+ I}] \times
{ 1\over ( 2\pi)^5 R V Q_5 } \int d^5 x Tr[ F_{-J} 
F_{-J }] 
\cr
=& { 2 g^2 \over R V Q_5 } {1\over r^2}
 \int dt \int dx^9 T_{++} \int dx^9  T_{--} \cr
=& { 2 g^2  E_L E_R \over R V Q_5 } {1\over r^2}\int dt 
}
}
where $R$ is the radius of the $9^{th}$ direction and $V =R_5R_6
R_7R_8$
is  the product of the radii in the other four  directions. 
$E_{L,R} $ are the left and right moving energies of the 
effective two dimensional theory  \actionfl .
Notice that we have calculated only the bosonic terms. Supersymmetry
then implies that the fermions appear in \potentialnew\ just
as another energy contribution.
The form of the operator in  \potentialnew\ is the identical to the
one that appeared in the calculation of the
fixed scalars greybody factors \cgkt .

\subsec{Calculation of the $v^2$ forces} 

If the probe is also moving, it will 
feel a force proportional to $v^2$ besides the static force. 
This is an effect which, in some sense, is of the same order of
magnitude as the static force. The static force is roughly equal
to the square of the brane excitation energy, while the $v^2$ force
will be proportional to the energy, we are free to vary the 
ratio of the energy above extremality and the velocity. 

Now we go back to the general case of $N+1$ parallel $p$ branes.
In order to calculate the $v^2$ force we have to compute a 
one loop diagram with two insertions corresponding to probe
fields and two insertions corresponding to the fields of the other
$N$ D-branes. We can calculate a general term involving 
the gauge field $F_1$ on the $U(N)$ piece and the gauge field
$F_2$ on the $U(1)$ part. We do the general calculation because
it is very simple.
Again,  we think in terms of string theory.
We take the gauge fields to be diagonal in group indices but
not necessarily commuting in the spacetime indices.
If we think in terms of the open string stretching between 
the two D-branes we see that the boundary conditions will be
\eqn\boundcond{\eqalign{
 \p_+ X^\mu (\sigma =0) =&
 \left ( { 1- F_1 \over 1 + F_1} \right )^\mu_{~~\nu} \p_-X^\nu(\sigma =0)~,
\cr
\p_+ X^\mu (\sigma =2\pi ) =&
 \left({ 1- F_2 \over 1 + F_2} \right)^\mu_{~~\nu} 
\p_-X^\nu(\sigma =2 \pi ) ~ ,
}}
where $F$ is regarded as a matrix with respect to the spacetime
indices
\callanconstant .

We see that this is equivalent to the boundary conditions
with 
\eqn\newbc{
F_1 \to \tilde F , ~~~~~~~~~~ F_2 \to 0~,
}
where 
\eqn\ftilde{
 \left({ 1- \tilde F \over 1 + \tilde F} \right)^\mu_{~~\nu} =
 \left({ 1- F_1 \over 1 + F_1} \right)^\mu_{~~\delta}
 \left({ 1+ F_2 \over 1 - F_2} \right)^\delta_{~~\nu}~~~.
}

In  the presence of $F_1$ and $F_2$ the effective action 
will be as in \result\ but in terms of $\tilde F$ as defined
by \ftilde .
Since all the lower order terms in $\tilde F$ vanish we only 
need to calculate  $\tilde F$ to first order in the field 
strength which gives
$\tilde F = F_1 - F_2 $.
Replacing this in \result we find
\eqn\resulttwo{\eqalign{
S_2 = & { c_{7-p} \over 8 ( 2\pi )^p  r^{7-p} } 
\int d^{1+p } x \ 4 \Tr [ F_{1 \mu \nu} F^{ \ \nu \rho}_1 ] 
Tr[F_{2 \rho \sigma }
F^{ \ \sigma \mu }_2]  + 2 Tr [ F_{1 \mu \nu} F_{1 \rho \sigma }]
 Tr [ F^{ \  \nu \rho }_2 F^{ \ \sigma  \mu }_2  ]
\cr
&  - {1 \over 2 } Tr [ F_{1 \mu \nu}F^{\  \mu \nu }_1 ] Tr[
F_{2 \rho \sigma }F^{\ \rho \sigma }_2 ]
-  Tr [ F_{1 \mu \nu}F_{ 1 \rho \sigma }  ] Tr[
F^{\ \mu \nu  }_2 F^{ \rho \sigma }_2 ]\cr 
& 
+ {\rm fermions} ~.
}}
We have written a separate trace over the group indices of the
probe to generalize the result to the case of
$U(N +M ) \to U(N) \times U(M)$ but in the discussion 
below we consider just $M=1$.

We will be giving  the probe some 
velocity
$F_{0i} = v^i $. In this case we see that
\resulttwo\ reduces to 
\eqn\twovel{
S_2 =  { c_{7-p}  \over   r^{7-p} } 
\int d^{1+p } x \ \left [ {v^2 \over 2 } \ g \ T_{00} 
- { 1 \over 2 (2\pi)^p }  Tr [ F_{1 i \mu } F^{ ~~\mu }_{1 j} ]
v^i v^j 
+ {\rm fermions} \right ] ~,
}
where 
\eqn\tzz{
T_{00} = { 1 \over g ( 2\pi)^p } \left\{ Tr [F_{1\mu 0} 
F_{1~ 0}^{~ \mu} ]
+ { 1\over 4 }  Tr [F_{1\mu \nu } F_{1}^{~ \mu \nu } ]   
+ {\rm fermions} \right\}
}
is the stress tensor associated to the unbroken $U(N)$ Yang-Mills 
theory.
Supersymmetry implies that the fermions will appear in \tzz\
contributing
to the stress tensor. 

We are interested  in the case where the velocity is
along the directions transverse to the branes. We mean that
the probe Wilson lines do not have a constant time dependence
($ F_{2 I 0} = 0$), in other words, there are no
winding fundamental strings dissolved on the probe. We are also interested
in evaluating \twovel\ in an average sense.
Under this conditions the term proportional to $v^iv^j$ in \twovel\
does not contribute since it will be proportional to $ D_\mu X^i D^\mu
X^j $, we can integrate by parts the covariant derivative and 
then the term vanishes using the equations of motion. 
This implies that \twovel\ becomes 
\eqn\twovelfin{
S_2 = { c_{7-p} g \over  r^{7-p} } E  \int dt  {v^2 \over 2 }~,
}
where $E$ is the total energy on the D-brane. 
This result if valid for all branes in all dimensions.
It includes 
as a particular case the one loop calculation of
\dps . In that case,
$p=5$ and  we have  an extremal state containing $Q_1$ instanton strings
and momentum $N$. The total  energy on the fivebrane worldvolume is 
\hms 
\eqn\energy{
E = {N\over R } + { R Q_1 \over g }~,
}
so that the probe metric becomes 
\eqn\twofivedim{
S= {R V  \over g} \int dt {v^2 \over 2 } \left( 1 + {g^2 N \over R^2 V r^2 }
+ { g Q_1 \over V r^2 } + o(g^2/ r^4)  \right)~,
}
where the $o(g^2/r^4)$ term should come from a two loop calculation
\dps .

\newsec{Supergravity calculation}

We start by writing the action of 
a test D-brane in the
background
of a near extremal black hole that carries D-brane charge \leigh
\eqn\class{
S  =  - { 1 \over g (2\pi)^p } \int d^{p+1} x \ 
e^{-\phi}\sqrt{ det G }  + A_{01...p }~,
}
where $G$ is the induced metric on the brane and $A_{p+1}$ is  
 the 
$p+1$ form potential that couples to the D-brane charge. 
The background is the corresponding supergravity solution
\hstro . 
For a solution carrying a single charge we find 
\eqn\class{ 
S = - { R^p \over g } \int dt  \left[  f^{-1} \sqrt{ h - f({\dot r}^2/h + r^2
\dot
\Omega^2 ) } + { \sinh 2 \alpha \over 2} 
{r_0^{7-p} \over r^{7-p} }f^{-1} \right] ~,
}
where 
\eqn\defg{
f = 1 + { r_0^{7-p} \sinh^2\alpha \over r^{7-p} } ~,~~~~~~~~~~~~~~
h = 1 - {  r_0^{7-p} \over r^{7-p} }~,
}
where $r_0$ and $\alpha$ are parameters related to the D-brane
charge, $N$,  and the mass above extremality, $E$, 
\eqn\param{
c_{7-p} N = { r_0^{7-p} \sinh 2 \alpha \over 2 g } ~,~~~~~~~~~~~~
E = { ( 9-p) R^p r_0^{7-p} \over 2 (7-p)  c_{7-p} g^2 }~,
}
with $c_{7-p}$ as in \defcp .
Expanding \class\ in powers of the velocity we find a static 
potential, a $v^2$ term, etc.
We start calculating the $v^2$ term
\eqn\vsquare{ { R^p \over g } \int dt {1\over 2} 
{ 1\over \sqrt{h}} \left( { \dot r^2 \over h } + r^2 \dot \Omega^2
\right)
}
We notice that 
the
coefficient is different for $\dot r^2 $ and $r^2 \dot \Omega^2$ while
in our result \twovelfin\ the coefficient was the same.
The resolution to this apparent contradiction is that the coordinate
$r$ of the spacetime solution is not necessarily the same as the 
coordinate $r$ of the D-brane calculation.
Let us define a new radial coordinate $\rho$ by the equation
\eqn\rhoeq{
{ d \rho \over \rho } = { d r \over r \sqrt{h} } ~ .
}
Using \rhoeq\ the  term \vsquare becomes 
proportional to $ v_\rho^2 = ( \dot \rho^2 + \rho^2\dot \Omega^2 )$,
so that 
$\rho$ is interpreted as the distance that appears in the
D-brane calculation.\foot{Note that the horizon at $r=r_0$ is not 
at $\rho =0$ as it  said in a previous version of this paper,
I thank E. Kiritsis for pointing that out.}
Expanding \class\ to leading order in $r_0^{7-p}/r^{7-p}$, taking 
into account the change of coordinates \rhoeq\ we get the 
velocity dependent term
\eqn\velocact{
S = { R^p \over g} \int dt { v^2_\rho \over 2 } ( 1 + { (9-p) \over
2 (7-p)} { r_0^{7-p} \over \rho^{7-p} } )~.
}
Expressing $r_0^{7-p}$ in terms of the energy \param\ we find
again
\twovelfin , in precise agreement with the D-brane probe calculation.
Notice that it was crucial to perform a change of
variables to find 
 agreement.
This quantity agrees for any $p$, this is due 
to the simplicity of the operator that couples to $v^2$: it is just
the
physical energy, so it is not renormalized by strong coupling effects
(large $g N$ effects).

Now let us turn to the static potential
\eqn\staticpot{
V =  { R^p \over g} f^{-1} \left[ \sqrt{h}  -1  + ( {\sinh 2\alpha \over
2 } - \sinh^2 \alpha ){r_0^{7-p} \over r^{7-p} } \right]~.
}
To leading order in $r_0^{7-p}/r^{7-p}$ {\it and } leading
order in $c_{7-p} g N/r^{7-p}$, (which means large distances)
we find
\eqn\class{ 
V= - { R^p \over g } { r_0^{7-p} \over 2 e^{2 \alpha} r^{7-p} 
}}
Using \param\ we see that \class\ is proportional to $E^2/N$ which
roughly agrees with \result .

We can also calculate the static potential for the case
of a fivebrane probe on a five dimensional near extremal black 
hole background carrying $Q_5$ D-fivebrane charge and $Q_1$ 
D-onebrane charge. We obtain the same formula for the
static potential as in \class\ (for $p=5$) but the 
formulas expressing the energies of the left and right
movers in terms of the parameter $r_0$ are different \hms\ and
lead  to \dps 
\eqn\classres{
S = - {2 g^2 E_L E_R \over R V Q_5  }{1\over r^2} ~,
}
in precise agreement with the the D-brane calculation \potentialnew .

\newsec{M(atrix) theory black holes}

According to the proposal of Banks, Fishler, Shenker and Susskind 
\bfss , 
 in order to calculate any process in M-theory we
should first boost the system along some direction so that
it carries a large longitudinal momentum. 
The D-brane theory of black holes can be naturally interpreted
in terms of M(atrix) theory \martinec \verlindebh .

The most natural black holes to consider are 
one dimensional extended solutions along the longitudinal dimension
which also carry longitudinal momentum.
They are homogeneous, translational invariant (but not boost invariant) 
 along the longitudinal
direction and localized in the transverse directions. 
An example is the extremal black hole solution corresponding
to zero branes lifted up to 11 dimensions.
 This solution is singular and should not
be taken too seriously very close to the horizon. 
We could, however, consider a near extremal solution 
which indeed has a well defined horizon and is nonsingular.
Alternatively we could compactify 5 or more dimensions and
get extremal black holes with nonzero horizon area 
by turning on some other charges.
As we boost along the longitudinal dimension the black hole
becomes closer to extremality.
In the finite $N$ proposal of Susskind 
\lennyfiniten ,
we can also have these momentum carrying black holes since 
they are supergravity solutions which are left invariant by 
translations along the null direction, so they are also 
solutions to M-theory compactified along a null direction.

We see that if $R$ is the radius of
the compactified direction we will have to 
take the limit
\eqn\limit{
  { r_g^{d-3} \over l_p^{d-3} } \sim N_h {l_p^2 \over R^2 } \to \infty 
}
in order to trust the  black hole  supergravity solution.
 We define $r_g$ to be the
typical
gravitational radius of the configuration, as the radius of the
d-2 transverse sphere were the redshift with the asymptotic observer 
becomes
of order one  (i.e. the metric becomes different by an order one
amount from the Minkowski metric).
$N_h$ is the total momentum carried by the black hole which
will be $N_h \le N$ than the total momentum of the system.

In \bfss  \oamb  \lifschytz \lima\  it was shown that
the gravitons indeed feel the right forces when they are in
the presence of some other graviton, or a two brane, etc.
It one had a system of zero branes (some  gravitons) which is
excited by some finite energy above extremality 
we expect that the system will stay for some time
in the ``stadium''  region of \dkpp \dkps\ where the noncommuting
properties of the matrices becomes important.
In the number of zero branes becomes large as in \limit\ then
we expect that the system will become a near extremal black hole.

A graviton propagating close to this near BPS configuration
will feel a force. 
The static force
is  proportional to the square of the energy of the 
excited system of gravitons in the light cone frame, this energy is
the same as the energy above extremality defined above.
If we have M theory on $T^p$ we have to consider 
a $p+1$ dimensional YM theory on the dual torus \bfss \taylor ,
 more
precisely, the corresponding theory coming form a nontrivial
fixed point. Since this calculation is IR dominated and finite in
the gauge theory,  we expect that the answer is independent of the
precise UV definition of the theory. 
For general $p$ there is agreement up to a numerical
coefficient for the leading power in $1/r$ of the force.
In the case of M(atrix) theory on $T^5$ 
and a configuration with $Q_1$ longitudinal fivebranes
the agreement is precise. This configuration is  a black
string in the six uncompactified dimensions. The description \verlindebh
\martinec\ is
almost the same as the description of the 5 dimensional black hole 
with $Q_5$ D-fivebranes and $Q_1$ D-fivebranes.
A difference is that in matrix theory we only have the gauge theory
on the branes, we do not have the infinite tower
of massive states.
The gauge theory 
calculation,
which in string theory is valid only for $ r \ll \sqrt{\alpha'}$,
 will
now have to be valid also for distances $r\gg \sqrt{\alpha'}$.

The agreement is more impressive for the $v^2$ force. This force
is universally proportional to the the energy. This calculation
can be viewed as a test of the equivalence principle in M(atrix)
theory.

We have concentrated in the one loop contribution, it would be 
interesting to study further corrections due to perturbative and
non-perturbative effects \mdns .

{\bf Acknoledgements}

I very gratefull to T. Banks, M. Douglas, N. Seiberg,
S. Shenker and A. Strominger 
for interesting  discussions. I also want to 
thank E. Kiritsis for pointing
an error.
This work was  supported in part by 
DOE grant
DE-FG02-96ER40559.

\listrefs

\bye